\input harvmac

\Title {UCD-PHY-95-11}
{{\vbox {\centerline{Supergeometry of     Three Dimensional Black Holes}}}}
  \bigskip
\centerline{Alan R. Steif}
 \bigskip\centerline{\it Department of
Physics}
\centerline {\it   University of California }
  \centerline{\it Davis, CA 95616 }
\centerline{\it steif@dirac.ucdavis.edu}
  \vskip .2in

\noindent
ABSTRACT:
We show how the supersymmetric properties of  three dimensional black holes can be obtained
algebraically.
 The  black hole solutions are constructed as quotients
of the supergroup $OSp(1|\,2;R)$ by a discrete subgroup of
its isometry supergroup. The
generators  of  the action of the isometry supergroup   which commute with these
identifications are found. These yield the supersymmetries
for the black hole as   found in recent studies as well as the
usual geometric isometries. It is  also shown that in the
limit of vanishing cosmological constant, the   black hole vacuum becomes
 a null orbifold,
 a solution     previously    discussed in the context of string theory.

\Date {}
\def\Ads3{three dimensional anti-deSitter space}
\def\ADS{ three  dimensional anti-deSitter space}
\def\sl{$SL(2,{  R})$}
\def\sll{SL(2,{  R})_L }

\def\so{$SO(2,1)$}
\def\osp{$OSp(1|\,2;R)$}
\def\a{\alpha}
\def\b{\beta}
\def\ra{\rightarrow}

\def\m{\mu}
\def\pa{\partial}

\def\th{\theta}

\def\l{\lambda}
\def\n{\nu}
\def\g{\gamma}
\def\pr{\prime}
\def\D{\bf d}
\def\d{\delta}

\def\D3{\hbox{$D_3\kern -10pt / \kern 10pt$}}
\def\Aslash{\hbox{$ A^a\kern -10pt / \kern +10pt$}}
\def\paslash{\hbox{$ \partial \kern -6pt / \kern +6pt$}}
 
\def\e{\epsilon}

\def\tphi{\tilde\phi}
\def\tt{\tilde t}
\def\tr{\tilde r}

\def\eps{\epsilon}

\newsec{Introduction}

Gravity in $2+1$ dimensions has no local dynamics.
Classical solutions to   the theory  in the absence of matter
are locally flat  {\ref\djt{S. Deser, R. Jackiw, and
 G. `t Hooft, {\it Ann. Phys.} {\bf 152} (1984) 220.}}, or     have constant
curvature, if a cosmological constant is present  {\ref\dj{ S. Deser and R.
Jackiw,
 {\it Ann. Phys.} {\bf 153} (1984) 405.}}.
 Nevertheless, non-trivial global effects are possible and can
yield  interesting  solutions  such as black holes  {\ref\btz{
M. Banados, C. Teitelboim, and J. Zanelli, {\it Phys. Rev. Lett.} {\bf 69}
(1992) 1849;  M. Banados, M. Henneaux,  C. Teitelboim, and J. Zanelli,
 {\it Phys. Rev. D} {\bf 48} (1993) 1506.}} {\ref\sc{For a review, see S. Carlip, 
``The $(2+1)$-Dimensional Black Hole'',  to appear in {\it Class. Quantum Gravity}.}}. A useful way of describing some vacuum
solutions to $2+1$ gravity is in terms of a quotient construction. One begins
with a simple symmetric space $\tilde {\cal S}$ and identifies points under the
action of  a discrete subgroup, $I$, of its isometry
group, $G$, to obtain a spacetime, ${\cal S}$. The fixed points of the group
action correspond to singularities of ${\cal S}$. The residual symmetry group
of the spacetime is the subgroup $H\subset G$ that commutes with $I$.

     In this paper, we study the  supergeometry of the  black hole
solutions.
 In Section 2, we review the $2+1$ dimensional black hole solutions
focusing attention on  their construction as quotients from the group manifold
{\sl}. We also discuss how in the limit of vanishing  cosmological constant,
the  $M=J=0$  black hole vacuum becomes the null orbifold of string theory.
 In Section 3, the solutions are imbedded in the  supergroup
{\osp}. The generators of the action of the isometry supergroup   which commute
with the black hole
identifications are found. The even generators yield Killing vectors. The
odd generators can be put into correspondence with two component spinors.
We obtain the same number of Killing vectors and spinors as found in studies
of their supersymmetric properties  {\ref\oh{O. Coussaert and M. Henneaux, {\it
Phys. Rev. Lett.} {\bf 72} (1994) 183.}}{\ref\jp{ J.M. Izquierdo and P.K.
Townsend, DAMTP preprint R/94/44,
gr-qc/9501018. }}  in the context of $2+1$  dimensional  anti-deSitter
supergravity {\ref\at{A. Achucarro and P. Townsend, {\it Phys. Lett. B}
 {\bf 180} (1986) 89.}}.


 \newsec{ 2+1 Dimensional Black Hole Solutions}

 $2+1$ dimensional black holes  {\btz} are   solutions to Einstein's equations
with a negative cosmological constant, $\Lambda$,
\eqn\einsteineqn{
G_{\m\n} +\Lambda g_{\m\n} = 0,\quad \Lambda <0
.}
 The metric for the   black hole solutions is  given by
\eqn\bhmetric{
ds^2  = - ({  r^2\over l^2} - M) d t^2 -Jdtd\phi+ ({  r^2\over l^2} - M  + 
{J^2\over 4r^2})^{-1} d  r^2 +
 r^2 d  \phi^2
, \quad   0 \leq \phi < 2\pi
}
where   $l \equiv (-\Lambda)^{-1/2}.${\footnote{*}{We  have set $G=1/8$.}}
 $M$ and $J$ are the mass and angular momentum. {\bhmetric} describes a black hole solution with outer and inner horizons
at $r =r_+$ and $r= r_-$ respectively where
\eqn\rpm{
r_{\pm }= l \bigl ( {M\over 2}\bigr )^{1/2}
\biggl ( 1\pm \biggl ( 1- \bigl ( {J\over Ml} \bigr )^2 \biggr )^{1/2}\biggr )^{1/2}
 .
}
The region $r_+  < r < M^{1/2} l$ defines an ergosphere, in which the
asymptotic timelike Killing field ${\pa \;\over \pa t}$ becomes spacelike.
  $M=J= 0$ is the black hole vacuum.
The solutions with  $- 1 < M < 0, \;J=0$  describe  point particle sources with   naked conical singularities   at
$r=0  $ {\dj}.
The    solution  with $M= -{1 },\; J=0 $ is anti-deSitter space.


 We now review    the construction of the black hole solutions 
as quotients of {\Ads3} {\btz}. It will be more useful for the later discussion to view {\Ads3} as the group manifold $SL(2,R)$ and the group of identifications
as a discrete subgroup of   $SL(2,R)_L \otimes SL(2,R)_R$, the isometry group of {\sl}.  
Every solution to {\einsteineqn} in $2+1$ dimensions   corresponds to
{\Ads3}
{\it locally}.  However, since one is  still free  to make discrete
identifications, the solution  can differ   {\it globally}.
 Three dimensional anti-deSitter space is most easily described in terms of
the three dimensional hypersurface
\eqn\ads{
-T^2 + X^2 - W^2 +  Y^2 = -l^2
}
imbedded in  the  four dimensional flat space with metric
\eqn\imbmet{
ds^2 = - d T^2 + d X^2 - d W^2 + d Y^2 .
 }
 The topology of {\ads} is $R^2\times S^1  $  with $S^1$ corresponding  to the
timelike circles $T^2 + W^2 = {  const. }$   Anti-deSitter space is the
covering space   obtained
by    unwinding  the circle.

The isometry group of {\ADS} is the subgroup of the  isometry group of the flat
space
{\imbmet} which  leaves {\ads} invariant. This is $SO(2,2)$ with rotations in
the $T-W$ plane which differ by $2\pi n$ {\it not} identified.
  The hypersurface  {\ads} describing       {\Ads3}  is the group manifold
of $SL(2,R)$ as can be seen from the imbedding
\eqn\gmat{
g ={1\over l}
\pmatrix{T + X   & Y - W \cr
               Y + W & T-X\cr}
  , \quad {\rm det}\, g =   (T^2 - X^2 + W^2 -  Y^2)/l^2 =1 .}
The metric {\imbmet} is  the bi-invariant metric
\eqn\binvariantmetric{
ds^2 = {l^2\over 2} {\rm Tr} (g^{-1}dg)^2 \;
.}
In this representation, the $SO(2,2)$ isometries  are  induced
by its two fold cover $SL(2,R)_L \otimes SL(2,R)_R$:
\eqn\action{
g\ra AgB,\,\, A, B \in SL(2,R) .
} It is a two-fold cover because
$(A,B)$ and $(-A, -B)$ induce the same element of $SO(2,2)$.
We can choose
\eqn\spgen{
L_3 = \pmatrix{1&0\cr
               0& -1 \cr
               },\quad
L_+ = \pmatrix{0&1 \cr
               0& 0 \cr
                },\quad
L_- = \pmatrix{0&0 \cr
               1& 0 \cr
                }\quad
}
 as generators of {\sl}.
The black hole solutions are obtained by identifying points in anti-deSitter space under the action
of a discrete subgroup of  $SL(2,R)_L \otimes SL(2,R)_R$.

\subsec{$M>0$: Black Hole Solutions}

The $M>0$ black hole solutions are obtained by  the identification
     {\btz} {\sc}
\eqn\bhiden{
g\sim A^ngB^n , \quad n \; {\rm integer}
}
where 
\eqn\bhI{\eqalign{ 
A  = \exp { (\pi{ (r_+ + r_-)\over l} L_3  )}  =
 &\pmatrix{e^{ \pi (r_+ + r_-)/l  }   &0\cr 0
&e^{- \pi (r_+ + r_-)/l } \cr},  \cr 
B  = \exp { (\pi{ (r_+ - r_-)\over l}L_3  )}  = 
&\pmatrix{ e^{ \pi (r_+ - r_-)/l  } &0\cr 0
& e^{- \pi (r_+ - r_-)/l  }\cr},  \cr
}}
 with  $L_3$  given in {\spgen} and $r_{\pm}$ given in {\rpm}.
This identification    is generated by 
\eqn\bhgen{
L =
{(r_+ + r_-)\over l}L_3^L +   {(r_+ - r_-)\over l}  L_3^R \in  sl(2,R)_L \oplus sl(2,R)_R, \quad \quad M>|J|/l 
.}
  For $J\neq 0$, $g$ has no fixed points under the action of  {\bhI}
consistent with the fact that  the rotating black hole solution
is non-singular. However, for the non-rotating $(J=0)$ black hole, there are fixed points      under the identification {\bhiden} 
which correspond to the singularity $ r =0$.

{\subsec{  Black Hole Vacuum }}

The $M=J =0$ black vacuum  is given by
\eqn\mzerometric{
ds^2  = - {  r^2\over l^2} d  t^2 + { l^2 \over  r^2 }
d r^2 +   r^2 d   \phi^2
, \quad 0 <  \phi < 2\pi  .
}
We first obtain its  $\Lambda \ra 0$ $(l\ra\infty)$ limit which  should
describe
 a locally flat metric.
Define the new coordinate
\eqn\vdef{
v  = 2t + 2l^2/r ,
}
which parameterizes outgoing null curves. The
  metric {\mzerometric} then becomes
\eqn\vac{
ds^2  = - {  r^2\over 4l^2} d  v^2  -  dvdr +  r^2 d   \phi^2
,   \quad 0 <  \phi < 2\pi .
}
 Now, as $l\ra\infty$, {\vac} has the smooth limit
\eqn\nullorb{
ds^2  =    -  dvdr +  r^2 d   \phi^2 ,\quad 0 <  \phi < 2\pi.
}
{\nullorb} is the metric for a {\it null orbifold} and  has been considered
previously in the context of string theory {\ref
\hs{G. T.  Horowitz and A. Steif,   {\it Phys. Lett. B}, {\bf 258}   (1991)
91.}}.   It has zero curvature and can be obtained by
identifying three-dimensional Minkowski space under the action of  a null boost.

  Like   the null
orbifold, the   black hole  vacuum can   be obtained by identifying
points  under the action of a null boost, but now in {\Ads3} rather than flat
space. 
Consider coordinates in {\Ads3} defined by the following imbedding
\eqn\vacimbed{
 \eqalign{
 U &\equiv T-  X= r\cr
V &\equiv T + X  =v -{rv^2\over 4 l^2} + r \phi^2 \cr
W &= {vr\over 2l} -l\cr
Y & = r\phi  . \cr     }}
  Translations   $(\phi\ra \phi +E )$
  correspond to null boosts  in $(U,V, Y)$ 
\eqn\nboost{\eqalign{
U &\ra U^{\pr} = U\cr
N_E:\quad\quad V&\ra V^{\pr} =V + 2EY + E^2U\cr
Y&\ra Y^{\pr} = Y+ EU \cr
W&\ra W^{\pr} = W . \cr
}}
$N_E$    can   be obtained  by a contraction, {\it i.e.} by 
 conjugating a Euclidean rotation of angle {$\theta$}  by a boost of velocity
$\beta $ in the simultaneous limit that
  $\beta  \ra 1$ and   $\theta\ra 0$  with $E= \theta / \sqrt{1-{\beta}^2 }$ held. 
   $r$ in {\vacimbed} labels the  $U= const. $ null surfaces  which $N_{E}$
leaves invariant.   Identifying
points under the action of
\eqn\vacI{
I = \{N_{2\pi n},\;   n\; {\rm integer}\}
}
corresponds to making $\phi$ periodic in $2\pi$.  Substituting {\vacimbed}
into {\imbmet}, we obtain   the   black hole vacuum {\vac}.
Translations in $v$ also preserve  the metric
{\vac} and correspond   to null boosts in the $(U, V, W)$ space with $Y$ fixed.
The set of fixed points of {\nboost}  are
\eqn\fixedpoints{
 {\cal L} = \{U=Y=0, W=-l\}
}
 and from {\vacimbed} is seen to correspond to the null singularity $r=0$.

 {}From {\nboost} {\gmat}, the   black hole vacuum is thus obtained by the identification
 \eqn\vacid{
g\sim A^n g B^{ n},\quad n \;{\rm integer}.}
where 
 \eqn\mzeroI{
  A = \exp { 2\pi   L_+}=  {\pmatrix{
1 & 2\pi  \cr
  0 & 1\cr }},
\quad   B = \exp { 2\pi   L_-}=  {\pmatrix{
1 & 0  \cr
  2\pi  & 1\cr }}
}
generated by 
\eqn\vacgen{
L_+^L + L_-^R \in   sl(2,R)_L \oplus sl(2,R)_R, \quad \quad {\rm Black \; Hole\; Vacuum} 
.}

\subsec{Extremal $M =|J|/l$ Solution}

In this section, we obtain the extremal $M =|J|/l$ solution 
as a quotient by a discrete subgroup of $ SL(2,R)_L \otimes SL(2,R)_R$.
We first review how the solutions are  constructed by identifying points in 
anti-deSitter space {\btz}.
Setting $M=J/l$ in {\bhmetric} yields the extremal solution
\eqn\extremal{
ds^2 =   - ({  r^2\over l^2} - M) d t^2 -Ml dtd\phi+{dr^2 \over  ({   r\over l } - {Ml\over 2r} )^2}   +  r^2 d  \phi^2
, \quad   0 <  \phi < 2\pi
}
The case  $M=-J/l$ can be obtained by  letting   $t\ra -t$.

It is useful to consider  Poincare coordinates {\ref\he{S. W. Hawking
and G.F.R. Ellis, {\it The Large Scale Structure of Space-time},
(Cambridge University Press, Cambridge, 1973).}} $(\l_+, \l_-, z)$ defined by the imbedding
\eqn\poincareimbedding{\eqalign{
T+X &= l/z\cr
T-X &  = l(z + ( \l_+  \l_- )/z  )\cr
W&= - {\l_+ - \l_- \over  2 z} l \cr
Y& = {\l_+  + \l_- \over  2 z} l   .\cr 
}}
Using {\imbmet},  the metric for anti-deSitter space in Poincare coordinates  takes
the form 
\eqn\adspoincare{
ds^2 = {l^2\over z^2} (d\l_+ d \l_-  + dz^2 )
.}
Consider  the one-parameter subgroup of $SO(2,2)$ transformations with parameter ${\chi}$  
 \eqn\translation{\eqalign{
\l_+ & \ra \l_+ + \chi\cr
\l_- &\ra  e^{(2M)^{1/2} \chi }\l_- + (2M)^{-1/2} ( e^{(2M)^{1/2}\chi} -1)\cr
z &\ra  e^{(M/2)^{1/2}\chi}z \cr 
}}
leaving {\adspoincare} invariant.
It was shown in {\btz} that   the extremal black hole {\extremal} is obtained by identifying   under {\translation} with $\chi= 2\pi$. {}  From {\poincareimbedding} and {\gmat}, 
the extremal black hole is obtained by the $SL(2,R)_L \otimes SL(2,R)_R$,
identification
 \eqn\extid{
g\sim A^n g B^{ n},\quad n \;{\rm integer}}
where   
 \eqn\extA{
A =    {\pmatrix{
e^{- (2M)^{1/2}\pi} & 0 \cr
 (2/M)^{1/2}\sinh{ ( (2M)^{1/2}\pi )} & e^{  (2M)^{1/2}\pi}\cr }},
\quad B =   {\pmatrix{
1 & 2\pi  \cr
  0 & 1\cr }}
  }
  generated by 
 \eqn\extremalgen{
L_-^L  - ({M\over 2})^{1/2} L_3^L + L_+^R \in  sl(2,R)_L \oplus sl(2,R)_R, \quad \quad {\rm\; Extremal \; Black\; Hole} 
.}

\subsec{$M<0,\; J=0 $ Solutions with Naked Singularities}

For the $M<0, J=0 $ solution, it is convenient to use   static coordinates  defined
by the imbedding
\eqn\staticcoor{\eqalign{
T &= \sqrt{\tr^2 + l^2 }\cos \, \tt/l,  \quad\quad\quad\quad \;\; W =
  \sqrt{\tr^2  +l^2 }\sin \, \tt/l, \quad  \;\;  \cr
 X &=    \tr \cos \, \tphi,  \quad\quad\quad\quad \quad\quad\quad\quad\;\; Y =
\tr \sin \, \tphi , \quad    \cr
}}
in terms of which the metric {\imbmet} for {\Ads3} takes the form
\eqn\staticmetric{
ds^2  = - ({\tr^2\over l^2} +1) d\tt^2 + ({\tr^2\over l^2} +1)^{-1} d\tr^2 +
\tr^2 d \tphi^2
   .
}
$\tt$ and $\tphi$ now parameterize rotations in the $T-W$ and $X-Y$ planes.
 The solution is now obtained by identifying
$\tphi$ periodically with period $2\pi \sqrt{|M|}.$    Rescaling the
coordinates
\eqn\rescaleneg{
\tr = r/\sqrt{|M|},\quad \tt = \sqrt{|M|} \,t, \quad \tphi =
\sqrt{|M|}\,\phi ,
}
one obtains   {\bhmetric} where
$\phi$   has   canonical
  period $2\pi $.

{}From {\gmat}, a rotation of angle $\theta$ in the $X-Y$ plane takes the form
{\action}
\eqn\slrotation{
\quad g \ra  \pmatrix{ \cos{\theta /2}
 & -\sin {\theta /2}\cr
\sin {\theta /2} &\cos{\theta /2} \cr }
g
\pmatrix{ \cos{\theta /2}
 &\sin {\theta /2}\cr
-\sin {\theta /2} &\cos{\theta /2} \cr }  .
}
Hence, the $M<0, J=0 $  solution is obtained by the 
 identification
\eqn\nakediden {
g \sim A^{-n}gA^n  
 } 
where
\eqn\nakedz{
 A = \exp { (\pi{\sqrt{|M|}}(L_+ - L_-))}   =
\pmatrix{ \cos{\pi \sqrt{|M|}}
 &\sin {\pi \sqrt{|M|}}\cr
-\sin {\pi \sqrt{ |M|}} &\cos{\pi \sqrt{ |M|}} \cr}
  .
} 
 generated by 
 \eqn\nakedgen{
L_+^L  -  L_-^L -  L_ +^R  + L_-^R
\in   sl(2,R)_L \oplus sl(2,R)_R, \quad \quad {\rm M<0,\; J=0  } 
.}

  The fixed points of the group action are $\{ X=Y=0\}$, and from {\staticcoor}
is seen to correspond to the singularity $r=0$.
These solutions are the anti-deSitter analog of the conical solution
{\djt} and were first constructed in {\dj}.

\newsec{Supergeometry}

In this section, we study the supergeometry of the black hole solutions. After
imbedding the black hole
spacetime in  the  supergroup {\osp}, one    finds the generators
of the isometry group of the supergroup which commute with the black
hole identifications. The even generators yield the usual
Killing vectors. However, in addition, there are odd generators of
the isometry group of  {\osp}
which are consistent with the black hole identifications.
 These can be
put into correspondence with two-component spinors. We find the same number of
these
Killing spinors as were found in studies of their supersymmetric properties
{\oh}{\jp}.  In {\oh},  it was pointed out  that the Killing spinors in the black hole
are those in anti-deSitter space which respect the identifications.
  Let us now  review the construction of  the supergroup {\osp}.

\subsec{ \osp }

Consider a Grassmann algebra, ${\cal A}$, generated by one Grassmann element,
$\e$
\eqn\algebra{
 {\cal A}= \{ z=a + b\e, \quad a,b \in R,\quad \e^2 =0\}
.}
$a$ and $b\e$ are the even and odd parts of $z.$
 {\osp}  is the set of linear transformations  of   $( \th^1, \th^2, x)$
of the form
\eqn\ospelt{
OSp(1|\,2;R)  = \biggl \{   M = \pmatrix{a&b&\a\cr
         c&d&\b\cr
         \g&\d&1\cr},\quad a,\ldots  \; {\rm even} ,\quad
 \a,\ldots\; {\rm odd} \biggr \}  }
  which preserve
\eqn\lineelt{
dl^2 = \eps_{ab} \th^a \th^b + x^2,\quad \eps_{12} = -\eps_{21} =1 
}
and where  $\th^1, \th^2$ are Grassmannian
satisfying
\eqn\thetas{
\th^1\th^1 =\th^2\th^2 =0,\,\,\{\th^1,\th^2\} =0,\,\, \{\epsilon, \theta^a\} =0
 .
}
The condition that $M$ preserves the line element  {\lineelt} implies the
relations
\eqn\cond{\eqalign{
ad-bc &=1 \cr
c\a -a\b &= -  \g \cr
d\a -b\b & = -  \d  .\cr
}}
Since these are three relations for 8 parameters, {\osp} is five dimensional.
{\osp} contains  {\sl} as a subgroup
\eqn\slsubgp{
   SL(2, R)\simeq  Sp(2,R)\simeq \biggl \{g = \pmatrix{a&b&0\cr
         c&d&0\cr
         0&0&1\cr},\;\;ad-bc=1\biggr\}  \subset OSp(1|\,2;R)  .   }

Consider   the following basis for the Lie algebra  $osp(1|2;R)$.
The even generators  are those in the ${sl(2,R)}$ subalgebra
and are given by {\spgen}
\eqn\evengen{
L_3 = \pmatrix{1&0&0\cr
               0& -1&0 \cr
             0&0&0\cr  },\quad
L_+ = \pmatrix{0&1 &0\cr
               0& 0&0 \cr
               0&0&0\cr },\quad
L_- = \pmatrix{0&0 &0\cr
               1& 0&0 \cr
               0&0&0\cr }\quad
}
  and the odd generators are
 \eqn\supergen{
Q_+ = \pmatrix{0&0&1\cr
               0& 0&0\cr
               0&-1&0\cr},\quad
Q_-= \pmatrix{0&0&0\cr
               0& 0&1\cr
               1&0&0\cr} .\quad
}
 They satisfy the algebra
\eqn\ospalgebra{
\eqalign{[L_3, L_+]& =  L_+,\quad [L_3, L_-]  =  -L_-,\quad
[L_+, L_-]  = L_3\cr
[L_3, Q_+] & = Q_+,\quad [L_+, Q_+]   = 0,\quad [L_-, Q_+]   = Q_- \cr
[L_3, Q_-] & = - Q_-,\quad [L_+, Q_-]   = Q_+ ,\quad [L_-, Q_-]   = 0 \cr
\{Q_+, Q_+\} & = -2L_+, \quad \{Q_-, Q_-\}    =  2L_-, \quad
\{Q_+, Q_-\}   =  L_3 .\cr
}}

  As we now show, the adjoint action of the {\sl} subgroup induces an $SO(2,1)$
transformation on the   $sl(2,R)$ subalgebra and an {\sl} transformation
on the odd generators $Q_{\pm}$. Consider the adjoint action by an element
\eqn\h{
h = {\pmatrix{ a & b&0\cr c&d&0\cr 0&0&1 \cr}} \in SL(2,R)
}
  on the Lie algebra $osp(1,2|R)$.
   On the  $sl(2,R)$ subalgebra, the  adjoint action
\eqn\adjointaction{
ad_h: L\ra h^{-1} L h
 }
induces the   transformation on  the basis {\evengen}
\eqn\basistransform{\eqalign{
L_3 &\ra (ad+bc) L_3 + 2bd L_+ - 2ac L_-\cr
L_+ & \ra cdL_3 + d^2 L_+ - c^2 L_-\cr
L_- & \ra -abL_3 -b^2 L_+ +a^2 L_- .\cr   }}
  This  is an {\so} transformation preserving
inner product 
\eqn\innerproduct{
<A,B> = {l^2\over 2} Tr (AB)   
}
 with  $h$ and $-h$ inducing the same element of
{\so}.
  Under the adjoint action {\adjointaction}, the odd generators {\supergen}
transform  as  \eqn\spinortransform{
Q_+\ra ad_h Q_+ = h^{-1} Q_+ h = dQ_+ -c Q_-,\quad
Q_-\ra ad_h Q_- = h^{-1} Q_-  h = -bQ_+ +a  Q_-
}
corresponding to the {\sl}transformation
\eqn\spintrans{
{\pmatrix{ Q_+\cr Q_-\cr}} \ra (h^{-1})^t {\pmatrix{ Q_+\cr Q_-\cr}}
.}
 
 A   vector on the $SL(2,R)$ submanifold of {\osp}
 at the point $g$ {\slsubgp}
  can be
 decomposed into a vector $w$ tangent to  {\sl}  and a transverse odd vector
 field $\psi$
  \eqn\decomposition{
 v= w+ \psi,\quad \psi  = \pmatrix{0&0&\a\cr
          0&0&\b\cr
          \g&\d&0\cr}
}
  with $\a,\ldots$ odd and satisfying {\cond}.
We associate with each odd vector field $\psi$ {\decomposition},
the spinor field
\eqn\spinorfield{
\psi  =  {\pmatrix {\bar a\cr \bar b\cr} } , \quad \a = \bar a \epsilon,\; \b
=\bar  b \epsilon .}

A right invariant basis of vector fields for {\osp}  on the 
{\sl} submanifold can be obtained by left
multiplication of   {\slsubgp} by the generators {\evengen} and {\supergen}.
The three vector fields obtained  from  {\evengen} are a right invariant basis
of vector fields
tangent to {\sl}
 while the two odd vectors obtained from
{\supergen}    yields the right invariant basis
of odd vector fields   given by
\eqn\rightbasis{
 \psi_+ = Q_+ g = \pmatrix{0&0&1\cr
         0&0&0\cr
         -c&-d&0\cr}
\quad \psi _- = Q_- g = \pmatrix{0&0&0\cr
         0&0&1\cr
         a&b&0\cr}
}
  with corresponding  spinors
\eqn\rightinvspinbasis{
\psi_+ ={\pmatrix {1\cr 0\cr} } \quad \psi_- =  {\pmatrix {0\cr 1 \cr}}
}
using {\decomposition} and {\spinorfield}.

\subsec{Supersymmetries}

Since  {\sl}  is a subgroup of   {\osp} , the black hole solutions which are
constructed as quotients of  {\sl}   can  also be viewed as quotients of  {\osp} . Since  {\osp} 
 is a group, its symmetry group with respect to a bi-invariant metric is ${OSp(1|\,2;R)_L}\otimes OSp(1|\,2;R)_R   $.
The symmetry group of the quotient $OSp(1|\,2;R)/ I$ is $H$ where
$H$ is the subgroup of  ${OSp(1|\,2;R)_L}\otimes OSp(1|\,2;R)_R   $
commuting with $I$
\eqn\hdef{
[H,I] =0,\quad H\subset OSp(1|\,2;R)_L\otimes OSp(1|\,2;R)_R   
.}
The even generators of  the Lie algebra of  $H$, $\cal H$, are the usual Killing symmetries while the odd generators
are the supersymmetries or Killing spinors. 
Given the odd generators, the corresponding Killing spinor fields
can be obtained by left or right multiplication by $g$ in {\slsubgp}. For the case of anti-deSitter space with
no quotient taken $(I=1)$, the full symmetry group is 
$H\in {OSp(1|\,2;R)_L}\otimes OSp(1|\,2;R)_R$
yielding $2\times 3 =6$  Killing vectors and $2\times 2 =4 $ supersymmetries.
Now we consider the black hole solutions. 

~From     {\ospalgebra},
we find that for the non-extremal black hole, there are two generators commuting with
$I$  {\bhgen}
\eqn\bhkilling{
 L_3^L ,  L_3^R \in {\cal H}    \quad  { \rm (Non-Extremal\;
Black \;Hole)}
}
implying there are
\eqn\mpos{
{\rm 2 \;Killing \; vectors\; and \; 0 \;Killing \;spinors }\quad {\rm (Non-Extremal\;
Black \;Hole)}
 .}
There are no Killing spinors because no non-trivial linear combination
of $Q_\pm$ commutes with $  L_3 $.

For the  black hole vacuum,    there are four generators commuting with $I$   {\vacgen}
\eqn\mzerokillingvectors{
L_+^L,
  L_- ^R , 
 Q_+^L,   Q_-^R ,\quad {\rm (Vacuum)}
}
implying
\eqn\mzerospinvec{
{\rm 2 \;Killing \; vectors\; and \; 2 \;Killing \;spinors }\quad ({\rm Vacuum})
 .}

{}For the extremal black hole solutions, using   {\ospalgebra}  we find  that
there are three  generators commuting with     {\extremalgen}
\eqn\extkillingvectors{
L_-^L - {(M/2)^{1/2}} L_3^L ,
 L_+^R,   Q_+^R , \quad {\rm\;( Extremal \; Black\; Hole)} 
}
implying
\eqn\extspinvec{
{\rm 2 \;Killing \; vectors\; and \; 1  \;Killing \;spinor },\quad {\rm\; (Extremal \; Black\; Hole)} 
  .}
  
 From      {\ospalgebra},
we find that for the $M<0$ solutions, there are two generators commuting with
$I$ {\nakedgen}
\eqn\bhkilling{
L_+^L  - L_-^L ,   L_+^R -L_-^R,\quad    (-1
<M<0,\; J=0)
}
implying
\eqn\mneg{
{\rm 2 \;Killing \; vectors\; and \; 0 \;Killing \;spinors }, \quad (-1<M<0, \; J=0)
 .}
There are no Killing spinors because no non-trivial linear combination of $Q_{\pm}$
  commutes with $L_+-L_-$.
For all the black hole solutions, the  two Killing vectors correspond  to
linear combinations of
  ${\pa\;\over \pa t}$ and  ${\pa\;\over \pa \phi}.$

We can also recover the Killing vectors and spinors for the self-dual
backgrounds considered in {\ref\ch{O. Coussaert and M. Henneaux, Universite
Libre de Bruxelles preprint ULB-TH 14/94, hep-th/9407181.}}.
The group of identifications   for a causally well-behaved
self-dual solution is a subgroup of one of the {\sl} factors, say {$\sll$},
generated by a spacelike generator. Since the left and right factors commute,
there
are two Killing spinors and three Killing vectors
coming from $OSp(1,2|R)_R$. From ${\sll}$, there are  zero Killing spinors and
 one  Killing vector.  Hence, for the self-dual solution there
are in total four Killing vectors and two Killing spinors.

 \vskip 20pt

\centerline{\bf Acknowledgements}

I  would like to thank Gary Gibbons and Paul Townsend
 for helpful discussions and Steve Carlip and  Yoav Peleg  for useful
comments on the paper.

 I  would also like to acknowledge the
 financial support of NSF grant NSF-PHY-93-57203 at Davis and the
SERC at Cambridge.

\baselineskip=30pt

\listrefs
\end